Susan Greenfield, *Mind Change: How Digital Technologies Are Leaving Their Mark On Our Brains*. New York: Random House, 2015. xvii + 348 pp. ISBN 9780812993820, $28.00 (hbk)

**Reviewed by:** Todd Davies, *Stanford University, USA*

Neuroscientist Susan Greenfield's *Mind Change* grew out of controversial statements she made as a member of the UK House of Lords. In a 2009 debate about websites, Greenfield recounts, “I decided to offer a perspective through the prism of neuroscience... the human brain adapts to the environment and the environment is changing in an unprecedented way, so the brain may also be changing in an unprecedented way” (p. xiii).

“Mind change” is Greenefield's umbrella term for digital technologies' effects, in a parallel to climate change. Greenfield's summary of the research is this: “social networking sites could worsen communication skills and reduce interpersonal empathy; personal identities might be constructed externally and refined to perfection with the approbation of an audience as priority, an approach more suggestive of performance art than of robust personal growth; obsessive gaming could lead to greater recklessness, a shorter attention span, and an increasingly aggressive disposition; heavy reliance on search engines and a preference for [Web] surfing rather than researching could result in agile mental processing at the expense of deep knowledge and understanding” (p. 265). To address these worries, she recommends that we all (1) deliberate about and decide “what kind of society we want, and what kind of individual traits we

This is a preprint of Todd Davies' review of Susan Greenfield's book *Mind Change: How Digital Technologies Are Leaving Their Mark on Our Brains*, first published on June 6, 2016, doi:10.1177/1461444816652614. The final, definitive version of this paper has been published in *New Media & Society*, 18(9): 2139-2141, October 2016 by SAGE Publications Ltd, value," (2) "take the pulse of societies around the world" by doing formal surveys of stakeholders, (3) do more research on technology's effects, and (4) invent "completely novel software that attempts to compensate for and offset any possible deficiencies arising from excessive screen-based existence" (pp. 269-270).

Greenfield's viewpoint is reflected, e.g., in studies showing that teachers have noticed declines in students' attention spans, and that they attribute this to digital technologies (p. 28). Her worries are shared by many media researchers and seem reasonable when she expresses them, in this book, as tentative and worthy of further study. But the book has attracted significant negative commentary in the British press and blogosphere, in part because of particular statements she made in the years prior to and surrounding its publication.

While the book has flaws, it is valuable as a public appeal to attend to new media's possible effects. Much of the research Greenfield discusses is not widely appreciated. She mentions, for example, Seltzer et al.'s (2012) finding that while teenagers' phone calls with parents led to oxytocin and cortisol levels similar to those during in-person interactions, their hormonal responses to text messaging were similar to teens "who did not interact with their parents at all" (pp. 130-131). Greenfield's inclusion of neuroscience findings sets this book apart from popular works drawn mostly from behavioral research. Among the peer-reviewed findings she cites are: enlargement in video gamers of an area in the nucleus accumbens associated with compulsive gambling (p. 42), dopamine release while playing a video game that appears comparable to using Ecstasy (pp. 157-158), and adolescent game addicts showing white matter abnormalities (p. 198).

The most notable criticism of this book has come from Bell et al. (2015), who focus on Greenfield's claim that screen media may be causing autism, and on her allegedly misleading portrayal of the evidence for other effects. Greenfield has been careless in public with the terms “autism” and “Autistic Spectrum Disorder”.[1] In this book, she attempts to distinguish between autism and “autistic-*like* traits, such as avoiding eye contact” (p. 136). But she maintains that early exposure to media might explain some of the rise in clinical autism (p. 137). On this latter point, she may be on shaky ground. The evidence does not appear to show technology effects early enough to cause autism. Greenfield's critics correctly say such statements could do more harm than good by stigmatizing parents. Regarding her use of the evidence about most other effects of online behavior, however, I find Bell et al.'s assessment overly harsh, *if* we judge *Mind Change* by standards applied to other popular books written by scientists, such as Pinker (2011). Greenfield acknowledges that digital technologies have benefits in many contexts. Her goal is to stir interest in the problematic effects that *might* be occurring. And she is able to call on neuroscience to bolster her worries. Known mechanisms of plasticity strongly predict that repeated experiences will have effects on the brain, but media neuroscience studies tend to be newer and less well established than the behavioral studies that are the focus of Bell et al.

1 See Bishop D V M (26 September 2014) Why most scientists don't take Susan Greenfield seriously. Retrieved from http://deevybee.blogspot.co.uk/2014/09/why-most-scientists-dont-take-susan.html

Greenfield's presentation of others' findings sometimes fails to paint a clear picture. Bell et al. point out that Greenfield does not clearly distinguish between effects of digital technology *use* and the abandonment of activities (e.g. children playing outdoors) that technologies displace: an important distinction for researchers and the public. In several cases, I found some contradictory results presented without explanation, or acknowledgement. We are told that paper books result in better reading comprehension than e-books (p. 216), and a few pages later, about a study showing them to be indistinguishable (p. 221). Chapters are generally presented as streams of results, which do not put findings into a framework.

I do not find the analogy with climate change compelling for effects of digital technologies, since media affect us more individually than greenhouse gases do. A better analogy might be processed food (p. 110), or even, in some cases, tobacco. Sana et al. (2013) found that students who saw others who were engaged in media multitasking performed more poorly themselves (Greenfield p. 218) – a media equivalent of the effects of secondhand smoke.

Online attacks on Greenfield have at times seemed personal. She has also been criticized for promoting her ideas publicly and not subjecting them to professional scrutiny. Competing philosophies are at play about the public role of scientists. Greenfield has chosen to shift her career toward popular writing, on topics on which she has not done original research, while trading on her status as a scientist. Should her book therefore be dismissed completely? No. Is this practice good for science and society, and do the circumstances in this case justify it?

Reasonable people could disagree.

**References**

Bell V, Bishop D V M, and Przbylski A K (2015) The debate over digital technology and young people. *BMJ* 351:h3064.

Pinker S (2011) *The Better Angels of Our Nature*. New York: Penguin Books.

Sana F, Weston T, and Cepeda N J (2013) Laptop multitasking hinders classroom learning for both users and nearby peers. *Computers & Education* 62, 24-31.

Seltzer L J, Prososki A R, Ziegler T E, and Pollak S D (2012) Instant messages vs. speech: Hormones and why we still need to hear each other. *Evolution and Human Behavior* 33, 42-45.